\documentstyle[preprint,aps,epsf]{revtex}

\begin{document}

\draft
\title{		Renormalization Group Flow in the Cutoff Yukawa Model 
			and a Scale Invariance in the Nambu-Jona-Lasinio Model}
\author{			Keiichi Akama}
\address{	Department of Physics, Saitama Medical College,
                     Kawakado, Moroyama, Saitama, 350-04, Japan}
\date{\today}
\maketitle

\begin{abstract}
We investigate the renormalization group flow 
	of the Yukawa model with a fixed momentum cutoff
	at the leading order in $1/N$,
	where $N$ is the number of the fermion species.
We demonstrate the scale invariance of  
	coupling constants of the Nambu-Jona-Lasinio model. 
\end{abstract}

\pacs{PACS; 11.10.Hi, 11.10.Gh, 11.15.Pg, 12.60.Rc }

The Nambu-Jona-Lasinio (NJL) model is 
	an excellent model which can render us a field theoretical 
	description of the quantum composite state,
	that is, a composite arising due to quantum effects \cite{NJL}. 
It is originally a model of composite mesons in analogy with 
	the superconductor theory, 
	and is also applied to the composite photon \cite{Bjorken}, 
	composite gauge bosons and Higgs scalars \cite{TCA,top},
	induced gravity \cite{indG}, and various collective modes 
	in nuclear and solid state physics. 
Though the model is not renormalizable,
	it is known to be equivalent to the renormalizable Yukawa model
	under the compositeness condition (CC) \cite{CC}.
Then the renormalization group (RG) approach may be a powerful tool 
	to investigate this model.
In fact many people attempted to study and apply the method \cite{RG,BHL}.
Most of them, however, considered the limit of the infinite momentum cutoff, 
	which necessitates some additional assumptions,
	such as existence of fixed point and ladder approximation.
In this letter, we take the momentum cutoff 
	as a large but finite physical parameter, 
	and make no further assumptions.
Then we investigate the RG flow 
	of the Yukawa model with a finite momentum cutoff
	at the leading order in $1/N$,
	where $N$ is the number of the fermion species.
It turns out that the NJL model with a finite momentum cutoff
	is entirely at the fixed point.
This illustrates the scale invariance 
	of the coupling constants
	in the NJL model.
This is important because it appears to contradict 
	with the widely adopted phenomenological use of the 
	running coupling constants in the NJL model \cite{BHL}.

We consider the renormalizable Yukawa model 
	for $N$ elementary fermions 
	$\psi _0=(\psi_{01},\psi_{02},\cdots,\psi_{0N})$ 
	and a elementary boson $\phi _0$
	with the following Lagrangian
\begin{eqnarray} 
	{\cal L}_{\rm Y}=\overline \psi _0i\!\not\!\partial \psi _0
	+g_0(\overline \psi _{\rm 0L}\phi _0\psi _{\rm 0R}+{\rm h.c.})
	+|\partial _\mu \phi _0|^2
	-m_0^2|\phi _0|^2
	-\lambda _0|\phi _0|^4		\label{LY}
\end{eqnarray} 
where $m_0$ is the bare mass of $\phi _0$, 
	$g_0$ and $\lambda _0$ are bare coupling constants, 
	and the subscripts ``L" and ``R" indicate chiralities.
Since the NJL model, which we want to consider in connection,  
	is not renormalizable,
	we introduce some regularization scheme with a finite cutoff.
We adopt the dimensional regularization where we consider everything 
	in $d(=4-2\varepsilon )$ dimensional spacetime 
	with small but non-vanishing $\varepsilon $.
By them we are not considering "the theory at the $d(=4-2\varepsilon )$",
	but that in $d=4$ with the momentum cutoff described by the scheme.
The parameter $\varepsilon $ roughly corresponds to $1/\log\Lambda $ 
	with the momentum cutoff $\Lambda $.
To absorb the divergences of the quantum loop diagrams due to (\ref{LY}), 
	we renormalize the fields, the mass, and the coupling constants as
\begin{eqnarray} 
&&	\psi _0= \sqrt {Z_\psi } \psi ,\ \ \   
	\phi _0= \sqrt {Z_\phi } \phi ,\ \ \   
	Z_\phi m_0^2= Z_m m^2, 
\\&&
	Z_\psi \sqrt {Z_\phi }g_0= {Z_g} g \mu ^\varepsilon ,\ \ \ \ \  
	Z_\phi ^2\lambda _0= Z_\lambda  \lambda  \mu ^{2\varepsilon },\ \ \ \ \  
\label{ZZ}
\end{eqnarray} 
where $\psi $, $\phi $, $m$, $g$, and $\lambda $ are the renormalized 
	fields, mass, and coupling constants, respectively, 
	$Z_\psi $, $Z_\phi $, $Z_m$, $Z_g$ and $Z_\lambda $ are
	the renormalization constants,
	and $\mu $ is a mass scale parameter 
	to make $g$ and $\lambda $ dimensionless. 
Then the Lagrangian ${\cal L}_{\rm Y}$ becomes
\begin{eqnarray} 
&&	{\cal L}_{\rm Y}=Z_\psi \overline \psi i\!\not\!\partial \psi 
	+Z_gg\mu ^\varepsilon (\overline \psi _{\rm L}\phi \psi _{\rm R}+{\rm h.c.})
\cr &&\ \ \ \ \ \ \ \ \ \ \  
	+Z_\phi |\partial _\mu \phi |^2
	-Z_m m^2|\phi |^2
	-Z_\lambda \lambda \mu ^{2\varepsilon }|\phi |^4.		\label{LYR}
\end{eqnarray} 
As the renormalization condition, 
	we adopt the minimal subtraction scheme, 
	where, as the divergent part
	to be absorbed into in the renormalization constants, 
	we retain all the negative power terms in the Laurent 
	series in $\varepsilon $ of the divergent (sub)diagrams.
Then the parameter $\mu $ is interpreted as the renormalization scale.
Since the coupling constants are dimensionless,
	the renormalization constants depend on $\mu $ 
	only through $g$ and $\lambda $, but do not explicitly depend on $\mu $.

Now we consider the NJL model in its simplest form given by the Lagrangian 
\begin{eqnarray} 
	{\cal L}_{\rm N}=\overline \Psi i\!\not\!\partial \Psi 
	+F|\overline \Psi _{\rm L}\Psi _{\rm R}|^2,
	\label{LNb}
\end{eqnarray} 
where $\Psi=(\Psi _{1}, \Psi _{2}, \cdots , \Psi _{N})$ 
	are $N$ fermions, 
	$F$ is a bare coupling constant.
The system is equivalent to that described by the 
	Lagrangian \cite{KK}
\begin{eqnarray} 
	{\cal L}'_{\rm N}=\overline \Psi i\!\not\!\partial \Psi 
	+(\overline \Psi _{\rm L}\Phi\Psi _{\rm R}+{\rm h.c.})
	-{1\over F}|\Phi |^2,
	\label{L'Nb}
\end{eqnarray} 
where $\Phi $ is an auxiliary scalar field.

Now we can see that the Lagrangian (\ref{LYR}) of the Yukawa model
	coincides with the Lagrangian (\ref{L'Nb}) of the NJL model, if 
\begin{eqnarray} 
	Z_\phi =Z_\lambda =0.	\label{CC}
\end{eqnarray} 
The condition (\ref{CC}) is the ``compositeness condition" (CC) \cite{CC}
	which imposes relations among the coupling constants $g$ and $\lambda $, 
	the mass $m$, and the cutoff parameter $\varepsilon $ in the Yukawa model 
	so that it reduces to the NJL model.
The perturbative calculation shows that 
	$g\rightarrow0$ and  $\lambda\rightarrow0$ 
	as $\varepsilon\rightarrow0$ at each order,
	and the theory becomes trivial free theory.
Therefore we fix the cutoff $\Lambda=\mu e^{1/\varepsilon}$ at some finite value.

We can read off from (\ref{L'Nb}) and (\ref{LYR}) 
	that the fields and parameters of the NJL and the Yukawa models
	should be connected by the relations
\begin{eqnarray} 
&&	\Psi=\sqrt{Z_\psi}\psi,\ \
	\Phi={Z_g g\mu^\varepsilon \over Z_\psi}\phi,\ \
	F= {\ Z_g^2 g^2\mu ^{2\varepsilon}
	/ Z_\psi^{2} Z_m m^{2}}.  
		\label{Cf}
\end{eqnarray} 
The last of (\ref{Cf}) is so-called ``gap equation" of the NJL model.
In terms of the bare parameters the CC (\ref{CC}) 
	corresponds to the limit
\begin{eqnarray} 
	g_0\rightarrow \infty , \ \ \ \ 
	\lambda _0/g_0^4\rightarrow 0.
\label{limbare}
\end{eqnarray} 
These behaviors may look singular at first sight,
	but they are of no harm 
	because they are unobservable bare quantities.

Thus the NJL model is equivalent to 
	the cutoff Yukawa model (i.e. the Yukawa model with a finite cutoff)
	under the CC (\ref{CC}).
Then the RG of the former 
	coincides with that of the latter under the condition (\ref{CC}).
Let us consider the latter (the cutoff Yukawa model)
	with special cares on the finite cutoff.
In our case, it amounts to fix $\varepsilon =(4-d)/2$ at some non-vanishing value.
The beta functions and the anomalous dimensions are defined as 
\begin{eqnarray} &&
	\beta _g^{(\varepsilon )}(g,\lambda )=\mu {\partial g\over \partial \mu }\ , \ \ \ \ 
	\beta _\lambda ^{(\varepsilon )}(g,\lambda )=\mu {\partial \lambda \over \partial \mu }\ ,\label{betadef}\\
&&
	\gamma _\phi ^{(\varepsilon )}(g,\lambda )={1\over 2}\mu {\partial \ln Z_\phi \over \partial \mu }\ , \ \ \ \ 
	\gamma _\psi ^{(\varepsilon )}(g,\lambda )={1\over 2}\mu {\partial \ln Z_\psi \over \partial \mu }\ ,
\end{eqnarray} 
where the differentiation $\partial /\partial \mu $ performed with
	$g_0$, $\lambda _0$, and $\varepsilon $ fixed.
Operating $\mu (\partial /\partial \mu )$ to the equations in (\ref{ZZ}) we obtain
\begin{eqnarray} 
	\left[ \beta _g^{(\varepsilon )}{\partial \over \partial g}+\beta _\lambda ^{(\varepsilon )}{\partial \over \partial \lambda }+\varepsilon \right] gJ=0, \ \ \ \ 
	\left[ \beta _g^{(\varepsilon )}{\partial \over \partial g}+\beta _\lambda ^{(\varepsilon )}{\partial \over \partial \lambda }+2\varepsilon \right] \lambda K=0,
\end{eqnarray} 
where $J=Z_g/(Z_\psi \sqrt {Z_\phi })$ and $K=Z_\lambda /Z_\phi ^2$.
Comparing the residues of the poles at $\varepsilon =0$, we obtain
\begin{eqnarray} 
	\beta _g^{(\varepsilon )}=-\varepsilon g+g{\cal D}J_1, \ \ \ \  
	\beta _\lambda ^{(\varepsilon )}=-2\varepsilon \lambda +\lambda {\cal D}K_1, \ \label{beta}
\end{eqnarray} 
where ${\cal D}=g(\partial /\partial g)+2\lambda (\partial /\partial \lambda )$, and 
	$J_1$ and $K_1$ are the residues of the simple poles 
	of $J$ and $K$, respectively.
On the other hand the anomalous dimensions are given by
\begin{eqnarray} 
	\gamma _\phi ^{(\varepsilon )}=-{1\over 2}{\cal D}Z_{\phi 1}, \ \ \ \ 
	\gamma _\psi ^{(\varepsilon )}=-{1\over 2}{\cal D}Z_{\psi 1}, \label{gamma}
\end{eqnarray} 
where $Z_{\phi 1}$ and $Z_{\psi 1}$ are the residues of the simple poles 
	of $Z_{\phi }$ and $Z_{\psi }$, respectively.
We can read off from (\ref{beta}) and (\ref{gamma})
	that $\beta ^{(\varepsilon )}$'s depend on the cutoff only through
	the first terms $-\varepsilon g$ and $-2\varepsilon \lambda $ of the expressions,
	while $\gamma ^{(\varepsilon )}$'s are independent of $\varepsilon $.
We should be careful not to neglect the cutoff dependence of $\beta ^{(\varepsilon )}$'s.

Explicit calculations at the one-loop level show
\begin{eqnarray} &&
	Z_\phi =1-{Ng^2\over 16\pi ^2\varepsilon }\ ,\ \ \ 
	Z_\lambda =1-{Ng^4\over 16\pi ^2\varepsilon \lambda }\ ,\ \ \ 
\label{ZZ1}
\\&&
	Z_g=Z_\psi =1-{g^2\over 16\pi ^2\varepsilon }.\ \ \ \
\label{ZZ1'}
\end{eqnarray} 
The CC $Z_\phi =Z_\lambda =0$ (eq.(\ref{CC})) with (\ref{ZZ1}) 
	connects the terms with different order in the coupling constants.
Accordingly, the expansion in the coupling constants fails 
	in the case of the NJL model.
Therefore we instead adopt the $1/N$ expansion by assigning 
\begin{eqnarray} 
	g^2\sim1/N,\ \ \ \lambda\sim1/ {N},
\label{gsim}
\end{eqnarray} 
which does not mix the different orders in CC (\ref{CC}) with (\ref{ZZ1}).
The expressions in (\ref{ZZ1}) contains all the contributions 
	of the leading order in $1/N$.
Applying (\ref{ZZ1}) -- (\ref{gsim}) to (\ref{beta}) and (\ref{gamma}), 
	we get, at the leading order in $1/N$,
\begin{eqnarray} &&
	\beta _g^{(\varepsilon )}=-\varepsilon g+{Ng^3\over 16\pi ^2}\ ,\ \ \ 
	\beta _\lambda ^{(\varepsilon )}=-2\varepsilon \lambda +{N(4g^2\lambda -2g^4)\over 16\pi ^2}\ ,\ \ \ \label{beta1}
\\&&
	\gamma _\phi ^{(\varepsilon )}={Ng^2\over 16\pi ^2}\ ,\ \ \ 
	\gamma _\psi ^{(\varepsilon )}=0.\ \ \ \label{gamma1} 
\end{eqnarray} 
The RG equation 
	with the functions in (\ref{beta1}) and (\ref{gamma1})
	determine the flow of the various quantities 
	with the increasing scale $\mu$.

The RG equations for the coupling constants $g$ and $\lambda $
	are given by (\ref{betadef}) with (\ref{beta1}),  
	and are solved as
\begin{eqnarray} 
		g^2={ 			1		
\over 
\displaystyle 		{N\over 16\pi ^2\varepsilon }+{\mu ^{2\varepsilon }\over g_0^2}	
}\ , \ \ \ \ \ \ 
		\lambda = {   
\displaystyle 	 {N\over16\pi^2\varepsilon }+{\lambda _0\mu ^{2\varepsilon }\over g_0^4}
\over 
\displaystyle 		\left( {N\over 16\pi ^2\varepsilon }+{\mu ^{2\varepsilon }\over g_0^2}\right) ^2 	
}\ ,
\label{sol1}
\end{eqnarray} 
where the integration constants have been determined 
	in accordance with (\ref{ZZ}).
We can confirm the results by deriving (\ref{sol1}) 
	directly from (\ref{ZZ}) with (\ref{ZZ1}).
In fact the RG flow of coupling constant 
	is entirely determined by (\ref{ZZ}).
In the infinite cutoff limit $\varepsilon \rightarrow 0$, 
	(\ref{sol1}) becomes
\begin{eqnarray} 
		g^2={		1            
\over 
\displaystyle 		a-{N\ln\mu ^2\over 16\pi ^2} 
}\ , \ \ \ \ \ \ 
		\lambda = {\displaystyle b-{N\ln\mu ^2\over 16\pi ^2}    
\over 
\displaystyle 		\left( a-{N\ln\mu ^2\over 16\pi ^2}\right) ^2 
}\ .
\label{sol0}
\end{eqnarray} 
	with $a=N/16\pi ^2\varepsilon +1/g_0^2$ 
	and $b=N/16\pi ^2\varepsilon +\lambda _0/g_0^4$ kept fixed.

Let us consider the properties of the NJL model 
	in the RG flow of the Yukawa model.
In the limit of NJL model (\ref{limbare}), the solution (\ref{sol1}) reduces to
\begin{eqnarray} 
	g^2={16\pi ^2\varepsilon \over N}\equiv g^2_{\rm NJL},\ \ \ \ 
	\lambda ={16\pi ^2\varepsilon \over N}\equiv\lambda_{\rm NJL}.
\label{CCsol1}
\end{eqnarray} 
This can also be derived by directly solving the compositeness condition 
	$Z_\phi =Z_\lambda =0$ in (\ref{CC}) with (\ref{ZZ1}).
The coupling constants (\ref{CCsol1}) for the NJL model
	are independent of the scale parameter $\mu$.
Namely, the NJL model is at the fixed point (\ref{CCsol1}) 
	in the renormalization flow of the Yukawa model.

\begin{figure}
\epsfysize=5cm\epsffile{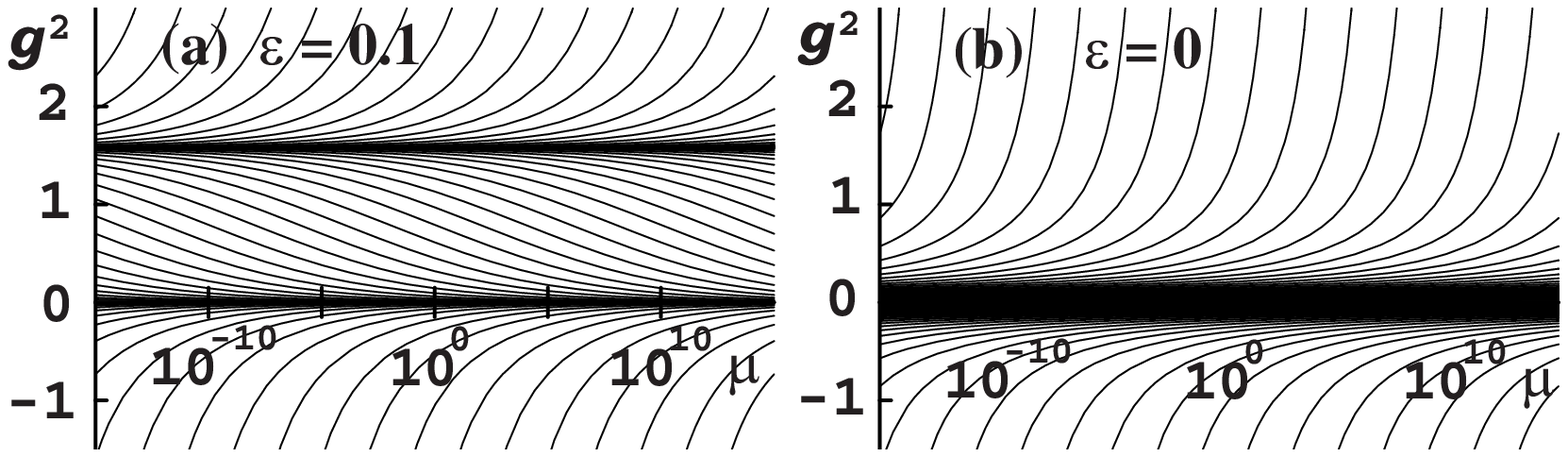}
\caption{ 
The RG flow of $g^2$ vs $\mu$ at 
(a) finite cutoff and (b) infinite cutoff.
}
\label{f1}
\vskip 1cm
\epsfysize=5cm\epsffile{ 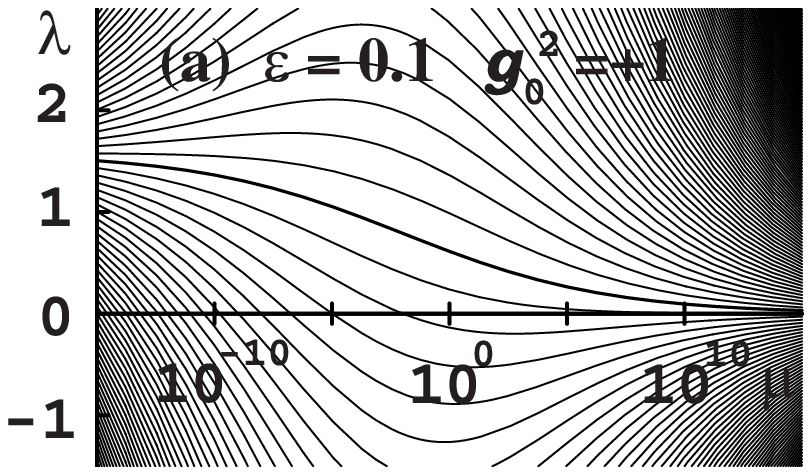}
\caption{
The RG flow of $\lambda$ vs $\mu$ for $g_0^2>0$ at (a) finite cutoff.
}
\label{f2}
\vskip 1cm
\epsfysize=5cm\epsffile{ 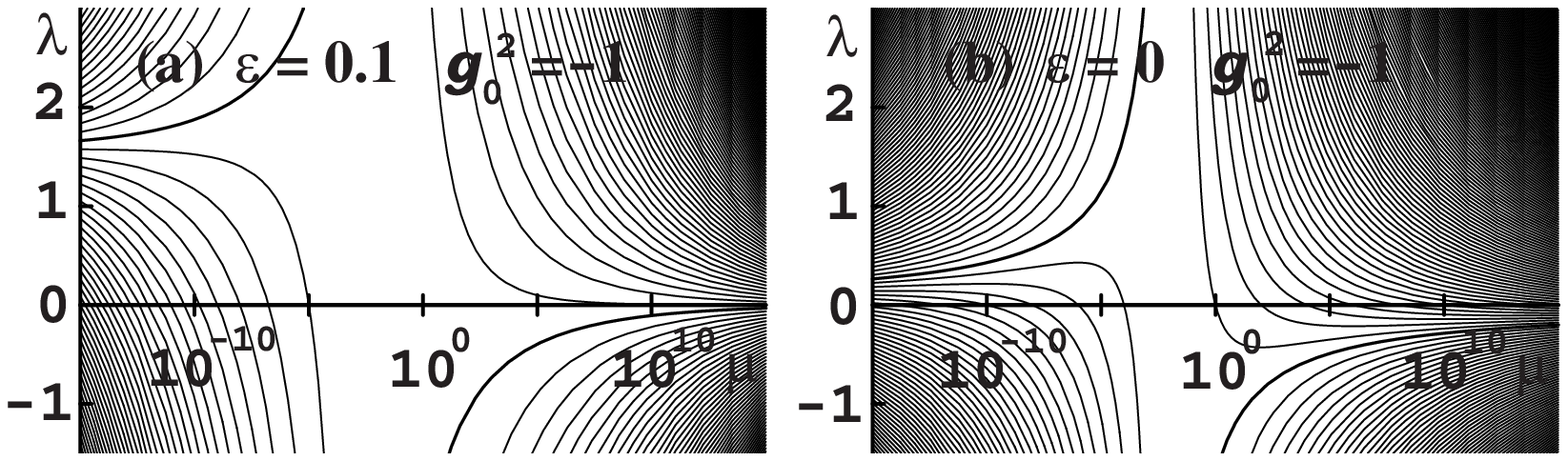}
\caption{
The RG flow of $\lambda$ vs $\mu$ for $g_0^2<0$ at 
(a) finite cutoff and (b) infinite cutoff.
}
\label{f3}
\vskip 1cm
\epsfysize=8cm\epsffile{ 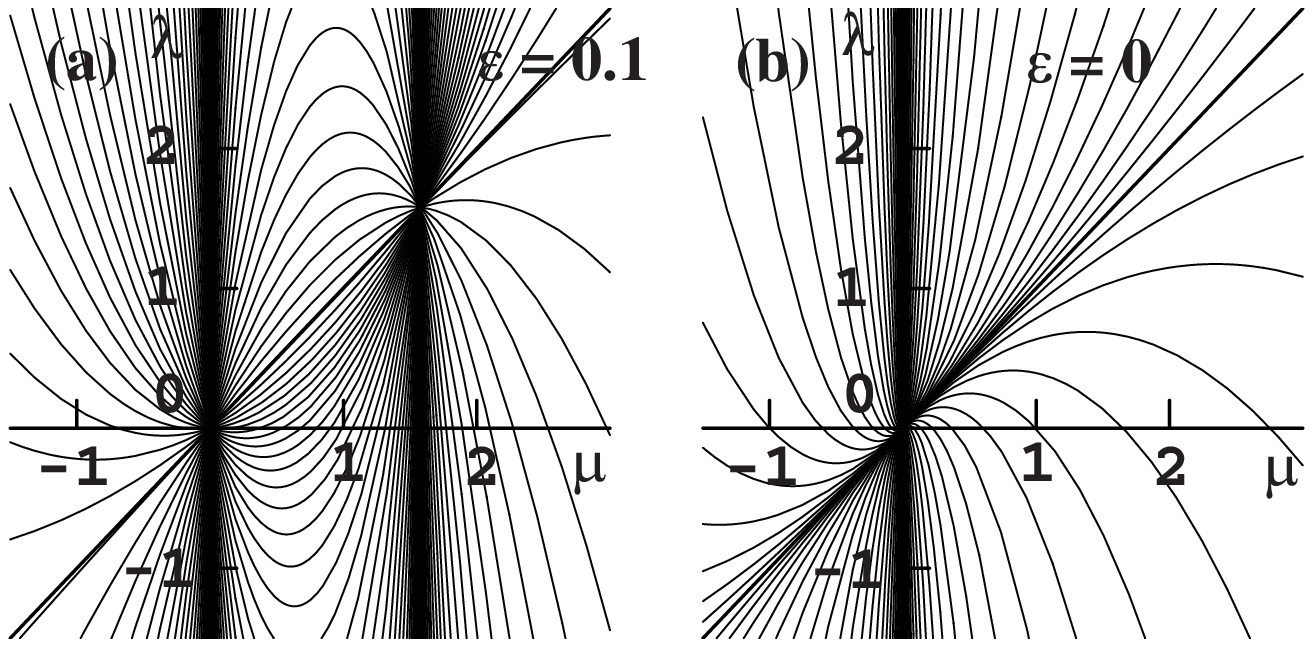}
\caption{
The RG flow in the $g^2$-$\lambda$ plane at 
(a) finite cutoff and (b) infinite cutoff.
}
\label{f4}
\end{figure}

In fig.\ \ref{f1}--\ref {f4}, we illustrate the typical RG flows 
	due to (\ref{sol1}) and (\ref{sol0}).
The number of the fermion species $N$ is typically taken as 10.
Fig.\ \ref{f1} shows the $\mu$-dependence of $g^2$
	for various values of $g_0^2$,
	while fig.\ \ref{f2} (fig.\ \ref{f3}) shows that of $\lambda$ 
	for various values of $\lambda_0$ 
	with a positive (negative) $g_0^2$, 
	typically taken as $ g_0^2=1$ ($ g_0^2=-1$).
Fig.\ \ref{f4} shows the flows in the $g^2$-$\lambda$-plane
	for various values of $\lambda_0$ 
	and $g_0^2$, 
The thick curves in fig.\ \ref{f2}--\ref{f4} are those with $g^2=\lambda$.
Fig.\ (a) of each figure shows the flow at finite cutoff $\varepsilon=0.1$,
	while fig.\ (b) shows its infinite cutoff limit $\varepsilon=0$.
If $g_0^2>0$, then $0<g^2<g^2_{\rm NJL}$,
	$g^2\rightarrow g^2_{\rm NJL}-0$, 
	and $\lambda\rightarrow \lambda _{\rm NJL}$
	as $\mu\rightarrow +0$,
	while $g^2\rightarrow +0$ and $\lambda\rightarrow 0$
	as $\mu\rightarrow \infty$.
If $g_0^2<0$ and 
$0<\mu<(-Ng_0^2/16\pi^2\varepsilon)^{1/2\varepsilon}\equiv\mu_{\rm c}$, 
	then $g^2>g^2_{\rm NJL}$,
	$g^2\rightarrow g^2_{\rm NJL}+0$, 
	and $\lambda\rightarrow \lambda _{\rm NJL}$
	as $\mu\rightarrow +0$,
	while $g^2\rightarrow \infty$ and $\lambda\rightarrow \pm\infty $
	as $\mu\rightarrow \mu_{\rm c}-0$.
If $g_0^2<0$ and $\mu>\mu_{\rm c}$, 
	then $g^2<0$, $g^2\rightarrow -\infty$ 
	and $\lambda\rightarrow \pm\infty $
	as $\mu\rightarrow \mu_{\rm c}+0$,
	while $g^2\rightarrow -0$ and $\lambda\rightarrow 0 $
	as $\mu\rightarrow \infty$.
Thus the $ (g^2,\lambda)=(g^2_{\rm NJL},\lambda_{\rm NJL})$ 
	is an infrared fixed point, 
	and $ (g^2,\lambda)=(0,0)$ is an ultraviolet fixed point.
The region $ g^2<g^2_{\rm NJL}$ is asymptotically free,
	though the region $ g^2<0$ is unphysical 
	because the Lagrangian is not hermitian.
On the other hand, in the region $g^2> g^2_{\rm NJL}$,
	$g^2$ and $|\lambda|$ blow up at $\mu=\mu_{\rm c}$,
	though it should not be taken serious,
	because the expansion fails in the region where $g^2$ is large.
In the infinite cutoff limit $\varepsilon\rightarrow0$,
	the fixed point $ (g^2,\lambda)=(g^2_{\rm NJL},\lambda_{\rm NJL})$ 
	moves to and fuses with the other fixed point $ (g^2,\lambda)=(0,0)$. 
Accordingly the physical asymptotic-freedom region 
	$0<g^2< g^2_{\rm NJL}$ disappears, leaving only the region
	where the coupling constants blow up.

Thus the NJL model is at a fixed point in the RG flow
	of the Yukawa model.
The coupling constants in the NJL model are scale-invariant,
	and does not run with the scale parameter.
Readers 	may have somewhat strange impressions to these statements,
	since many phenomenological models in literature uses 
	'running coupling constants in NJL type models' \cite{BHL}.
The discrepancies may depends on the details what they exactly mean
	by the running, the NJL model etc..
We can, however, trace back the reason of scale invariance
	to the fact that beta functions vanish 
	due to the compositeness condition.
In fact if we substitute the solution (\ref{CCsol1}) 
	of the compositeness condition,
	the beta functions (\ref{beta1}) vanish.
It is further traced back to the fact that the scale invariance of
	the relation  (\ref{ZZ}) under the compositeness condition (\ref{CC}).
Thus we expect that the scale invariance holds not only in 
	the leading order in $1/N$, but also in all order. 
A formal proof of this statement will be given in a separate paper \cite{beta}.

\end{document}